\documentclass[%
 reprint,
groupedaddress,
nofootinbib,
 amsmath,amssymb,
 aps,
pra,
]{revtex4-2}

\usepackage[utf8]{inputenc}
\usepackage[english]{babel}

\usepackage{amsmath}
\usepackage{amssymb}
\usepackage{bbm}
\usepackage{bm}
\usepackage{braket}
\usepackage{graphicx}
\usepackage{overpic}
\usepackage{dsfont}

\usepackage{siunitx}

\usepackage[caption=false]{subfig}

\usepackage{appendix}

\usepackage[usenames,dvipsnames]{xcolor}
\usepackage[colorlinks=true,citecolor=MidnightBlue,linkcolor=MidnightBlue,urlcolor=MidnightBlue]{hyperref}
\usepackage[capitalise]{cleveref}
\usepackage{soul}

\def\<{\langle}
\def\>{\rangle}

\def\ha{{\hat{a}}}
\def\hb{{\hat{b}}}

\def\hs{{\hat{s}}}

\newcommand{\nn}{\nonumber}

\begin{document}

\title{Linear Optical Schemes to Postselect High-Dimensional Dicke States}

\author{Daniel Bhatti$^{1}$, William J. Munro$^{2}$, Seungbeom Chin$^{2}$}
\affiliation{
$^1$Networked Quantum Devices Unit, Okinawa Institute of Science and Technology Graduate University, Okinawa 904-0495, Japan \\
$^2$Quantum Engineering and Design Unit, Okinawa Institute of Science and Technology Graduate University, Okinawa 904-0495, Japan
}

\begin{abstract}
Multipartite entanglement is an essential quantum resource for various distributed quantum applications. One promising method for preparing multipartite entanglement is to interfere independent photons at linear optical interference setups. While heralding the successful interference and thereby the state generation is often costly, postselecting entangled states provides an achievable alternative in this framework. We introduce a family of interference schemes for postselecting symmetric qudit Dicke states, useful resources in quantum communication and variational quantum computing. We present schemes with and without ancillary photons and show that using ancillary photons can exceed the upper bound on the success probability of schemes without ancillary photons. Our results accommodate a wide range of linear optical schemes, providing multiple viable approaches for postselecting Dicke states.
\end{abstract}

\maketitle

\section{Introduction}

High-dimensional entanglement offers multiple advantages over two-dimensional entanglement, including denser information encoding, increased noise tolerance, and simpler experimental setups~\cite{Wang2020,Erhard2020,Friis2019,Goel2024}. Especially photonic setups would benefit from the generation of high-dimensional entanglement and the accompanying reduction of photons, since photon transmission is strongly affected by noise and loss. However, as in two dimensions, it is important to choose the entangled states based on the task, the generation cost, and their noise and loss properties. While GHZ states possess maximal quantum correlations among all their qubits~\cite{Murta2020} and efficient schemes for generating them exist~\cite{Chin2024,Bhatti2025}, they maximally suffer from photon loss~\cite{Murta2020}. This motivates the generation of alternative high-dimensional multipartite entangled states that retain strong entanglement while improving robustness and scalability.

A class of more tolerant entangled states against photon loss is the Dicke states~\cite{Dicke1954, Kiesel2007}.
Symmetric $N$-qudit Dicke states consist of $k_{j}$ qudits per qudit state $\ket{j}$, with $\sum_{j}k_{j}=N$, and are defined as the sum over all symmetric permutations of the $N$ qudits~\cite{Wei2003}, i.e.,
\begin{align}
\label{eq:DickeStateQudit}
    & \ket{D_{N}^{\{k\}}} \nonumber \\ 
    & = \sqrt{\frac{\prod_{j} k_{j}!}{N!}} \sum_{\sigma \in S_{N}^{\{k\}}} \ket{\underbrace{0\ldots 0}_{k_{0}}\underbrace{1\ldots1}_{k_{1}}\ldots\underbrace{(d-1)\ldots (d-1)}_{k_{d-1}}},
\end{align}
where $S_{N}^{\{k\}}$ denotes the symmetric group that permutes the order of qudit states without multiple counting.
 
Dicke states have been shown to be useful in different applications. First, Dicke states are a natural benchmark for studying entanglement under particle loss, since their permutation symmetry and fixed-excitation structure make them highly robust multipartite states~\cite{Kiesel2007, Zhang2025}.
Furthermore, Dicke states are useful for quantum communication and secret sharing schemes~\cite{Wang2017, Lipinska2018, Grasselli2019, Li2024}, and have advantages in quantum repeater and network schemes~\cite{Prevedel2009, Miguel-Ramiro2023, Roga2023, Illiano2024}.
Recently, Dicke states have received attention in the context of variational quantum computing~\cite{Brandhofer2022}.

A feasible approach to make entangled multiphoton states available in optical experiments is postselection. So far, postselection has led to experimental realizations of symmetric qubit Dicke states with up to six photons~\cite{Prevedel2009,wieczorek2009experimental, Chen2023}.
In theory, more general schemes for postselecting~\cite{Kiesel2010, Maser2010, Kasture2018, Gu2019, Zhu2020, Chen2023, Lim2005, Kim2020, chin2021graph} and also heralding~\cite{kang2025heralded} photonic qubit Dicke states exist.
However, a thorough discussion of linear optical setups for postselecting multiphoton qudit Dicke states is still missing.

In this work, we address this open problem by introducing linear optical schemes for postselecting qudit Dicke states using independently prepared single-photon resources. We present protocols both with and without ancillary photons, revealing a trade-off between experimental simplicity and enhanced success probability. Together, these schemes encompass a range of resource requirements and interferometric structures, offering practical routes for generating high-dimensional Dicke states in photonic platforms depending on the desired balance between resource overhead and performance.

The paper is organized as follows.
\cref{sec:Operator-level} discusses an operator-level scheme for generating qudit Dicke states that captures the essential structure of linear optical circuits, thereby becoming a blueprint for our approach. 
In  \cref{sec:SchemesQudits}, we present actual linear optical circuits for generating $N$-partite qudit Dicke states with $N$ photons, i.e., without ancillary photons.
In \cref{sec:SchemesQuditsAncilla}, we discuss schemes with ancillary photons and show that such schemes can increase the success probabilities.
In \cref{sec:SpecialCases}, we discuss our results for two different cases, i.e., qubits and qutrits, and compare the success probabilities of the schemes with and without ancillary photons.
\cref{sec:conclusion} provides concluding remarks and discussions.

\section{Operator-level approach to generate multipartite entanglement}
\label{sec:Operator-level}

Our goal is to design linear optical schemes for postselecting symmetric Dicke states. As a first step, we analyze the problem at the \emph{operator level}, without reference to explicit optical components. For this, we briefly review how to transform quantum states linearly by mapping input to output operators, as known from linear optical quantum computation~\cite{knill2001scheme, Scheel2003}. This abstract formulation captures the essential structure of the transformations that generate entanglement through a single linear operator. In the subsequent sections, we show how these operators can be implemented in linear optical networks, providing explicit experimental schemes.
We consider quantum systems where single identical particles (in principle, they can be either bosons or fermions; however, in this work, we only consider photons) 
evolve linearly with the particle number preserved. While the most general form of transformation operators that rotate spatial modes and internal states is explained in Ref.~\cite{chin2021graph}, we restrict our case to the transformation operators that only affect the spatial distributions of particles.

Suppose we have $M$ particles with a $d$-dimensional internal degree of freedom in a set $a$ of $L$ spatial modes. 
In the second quantization language, the creation operator in the $j$th spatial mode is denoted as $\ha^\dagger_{j,s_j}$, where $s_j$ denotes the $d$-dimensional internal degree of freedom (e.g., multi-rail position of photons in optical systems). Since the qudit information is encoded with $s_j$, we set $s_j \in \{0,1,\cdots, d-1\}$.
The particle evolves under a linear transformation operator $T$ and arrives at another set $b$ of $L$ spatial modes. Then, the transformation of the single-particle state is described as
\begin{align}\label{eq:linear_evolution}
 		\ha_{j,s_j}^\dagger \to
		\sum_{l=1}^L T_{jl}\hat{b}_{l,s_j}^\dagger,
\end{align} where $T_{jl}~( \in \mathbb{C})$ satisfies $\sum_{l}|T_{jl}|^2 =1$ and $\hat{b}_{l,s_j}^\dagger$ denotes the creation operator at the second set of spatial modes. We can see that the internal state of the particles is preserved as $s_j$ under the transformation. 

The total $M$-particle state transformation is given by
	\begin{align}\label{eq:N_transf}
		\prod_{j=1}^L\ha_{j ,s_j}^\dagger|\text{vac}\>\to \prod_{j=1}^L \Big(\sum_{l=1}^L T_{jl}\hat{b}_{l ,s_j }^\dagger\Big)|\text{vac}\>.	
	\end{align}
In this paper, we postselect states with fixed numbers in each mode to generate multipartite entanglement.
The superposition of such states determines which type of entangled state can be postselected. 

\section{Postselecting High-Dimensional Dicke States without ancillary particles}
\label{sec:SchemesQudits}

In this section, we discuss schemes for postselecting symmetric qudit Dicke states without ancillary particles, i.e., postselecting $|D_N^{\{k\}}\>$ using $N$ photons. We first introduce an operator-level scheme by suggesting suitable $T_{jl}$ in Eq.~\eqref{eq:N_transf} to generate our target state. For realizing these states, one must choose photonic degrees of freedom capable of implementing photonic qudit states $\ket{j}$, with $j=0,1,\ldots,d-1$, e.g., path, frequency, time-bin, or orbital angular momentum~\cite{Erhard2020, Chin2024}.

\subsection{Operator-level scheme}

Initially, we  prepare $N$ particles, of which $k_j$ particles have the internal state $|j\>$ ($j\in \{0,1,\cdots, d-1\}$):
\begin{align}\label{eq:initial_no_ancillar}
    &|\Psi_{\text{in}}\> \nonumber \\
    &{}={} \prod_{l_0=1}^{k_0}\ha^\dagger_{l_0,0}\prod_{l_1=k_0+1}^{k_0+k_1}\ha^\dagger_{l_1,1}\cdots \prod_{l_{d-1}=\sum_{h=0}^{d-2}k_{h}+1}^{N}  \ha^\dagger_{l_{d-1},d-1}|\text{vac}\> .
\end{align}
A linear transformation operator $\tilde{T}$ that generates $|D_N^{\{k\}}\>$ is given by $\tilde{T}_{jl} = \frac{1}{\sqrt{N}}$ for all $j,l \in \{1,2,\cdots, N \}$, i.e., 
\begin{align}\label{no_ancilla_T}
\tilde{T}=
\frac{1}{\sqrt{N}}
\begin{pmatrix}
    1 & 1 & \cdots & 1 \\
    1 & 1 & \cdots & 1 \\
    \vdots & \vdots & \ddots & 1 \\
    1 & 1 & \cdots & 1 \\     
\end{pmatrix},
\end{align}
which was suggested, e.g., in Refs.~\cite{Maser2010,Kasture2018,chin2021graph}. Note that each particle is split into all the spatial modes with the same probability amplitudes, hence preserving the permutation symmetry of the Dicke state.

\begin{figure}
    \centering
    \includegraphics[width=1\linewidth]{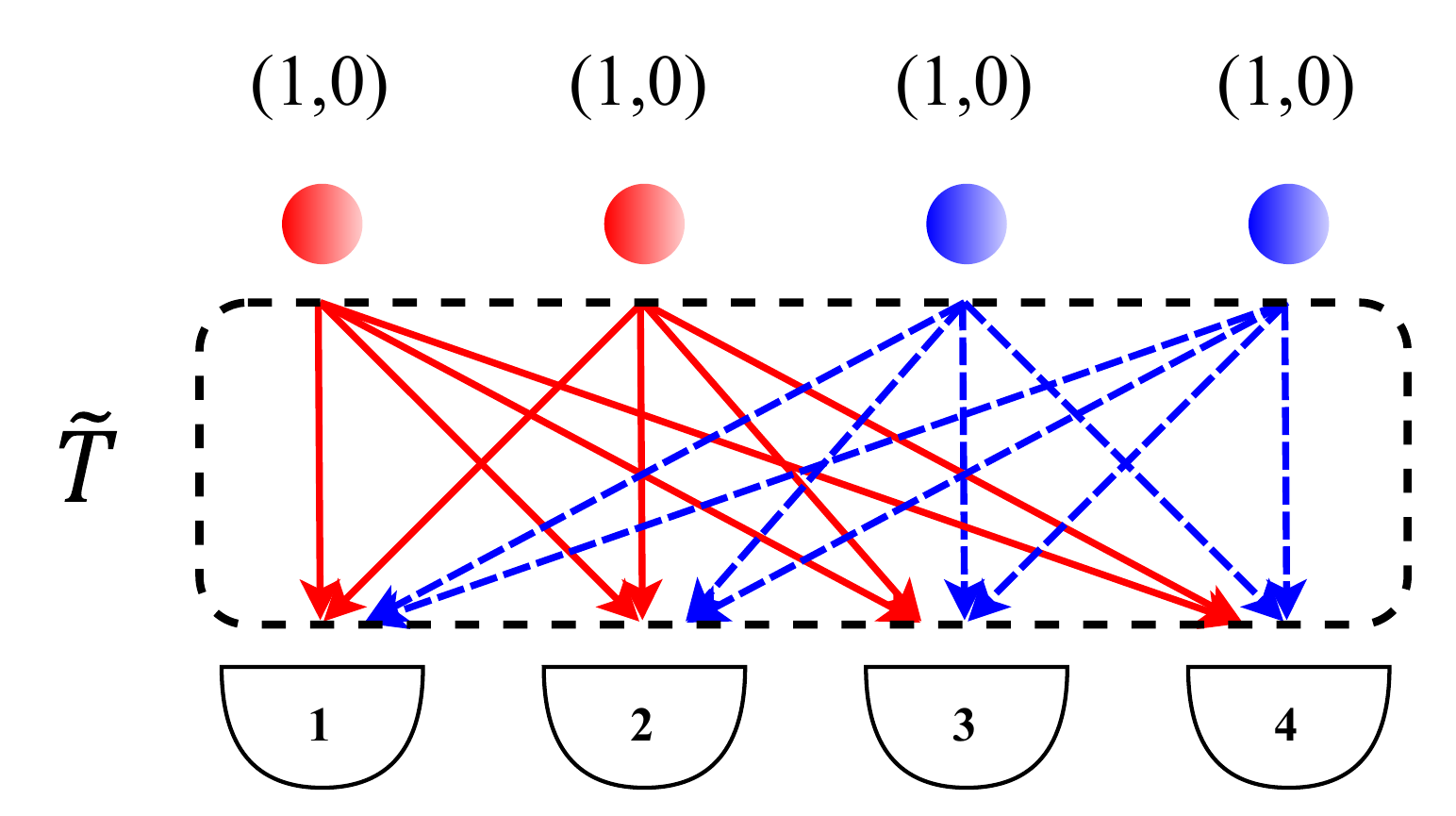}
    \caption{A linear operator that generates $|D_4^{\{2,2\}}\>$ with four particles. Two red particles in the first and second modes are in the internal state $|0\>$ and the other blue ones in $|1\>$. All the particles are spread into all four output modes by the transformation operator~$T$.}
\label{fig:linear_transformation_no_ancillar}
\end{figure}

Then the initial state~\eqref{eq:initial_no_ancillar} is transformed by $\tilde{T}$ into
\begin{align}
\label{qubit_operator_N} 
& \frac{1}{\sqrt{\prod_{j=0}^{d-1}k_j!}\sqrt{N}^N} \nn \\
& \times \Bigg( \sum_{m_0=1}^N \hat{b}^\dagger_{m_0,0} \Bigg)^{k_0}\cdots \Bigg( \sum_{m_{d-1}=1}^{N} \hat{b}^\dagger_{m_{d-1},d-1} \Bigg)^{k_{d-1}}|\text{vac}\>  ,
\end{align}
where $\frac{1}{\sqrt{\prod_{j=0}^{d-1}k_j!}}$ is multiplied as a normalization factor. 
We postselect the cases in which no particles arrive in the same mode.
In total, the postselection yields $N!/\prod_{j=0}^{d-1}k_j!$ distinct terms, each $\prod_{j=0}^{d-1}k_j!$ times. These terms correspond to all possible permutations of the $N$ qudits, and the final state is given by 
\begin{align}
\label{eq:NormalizedDicke}
\sqrt{\frac{N!}{N^N}}|D_N^{\{k\}}\> . 
\end{align} 
We can easily check that the postselected final state is indeed the Dicke state using the linear quantum graph picture
introduced in Ref.~\cite{chin2021graph,chin2024shortcut}. 
For a simple example, when $(N,k)=(4,2)$ (see Fig.~\ref{fig:linear_transformation_no_ancillar}), Eq.~\eqref{qubit_operator_N} becomes
\begin{align}
 \frac{1}{\sqrt{2\cdot 2}\sqrt{4}^4}&(\hat{b}^\dagger_{1,0} + \hat{b}^\dagger_{2,0} +\hat{b}^\dagger_{3,0}+\hat{b}^\dagger_{4,0})^2\\ \nn 
 &\times (\hat{b}^\dagger_{1,1} + \hat{b}^\dagger_{2,1} +\hat{b}^\dagger_{3,1}+\hat{b}^\dagger_{4,1})^2|\text{vac}\rangle.
\end{align}
After the postselection, the final state is given by
\begin{align}\label{operator_final}
  &  \frac{2\cdot 2}{\sqrt{4}\sqrt{4}^4}(\hat{b}^\dagger_{1,0}\hat{b}^\dagger_{2,0}\hat{b}^\dagger_{3,1}\hat{b}^\dagger_{4,1} + \hat{b}^\dagger_{1,0}\hat{b}^\dagger_{2,1}\hat{b}^\dagger_{3,0}\hat{b}^\dagger_{4,1}+\hat{b}^\dagger_{1,0}\hat{b}^\dagger_{2,1}\hat{b}^\dagger_{3,1}\hat{b}^\dagger_{4,0} \nonumber \\
&+\hat{b}^\dagger_{1,1}\hat{b}^\dagger_{2,0}\hat{b}^\dagger_{3,0}\hat{b}^\dagger_{4,1} +\hat{b}^\dagger_{1,1}\hat{b}^\dagger_{2,0}\hat{b}^\dagger_{3,1}\hat{b}^\dagger_{4,0} + \hat{b}^\dagger_{1,1}\hat{b}^\dagger_{2,1}\hat{b}^\dagger_{3,0}\hat{b}^\dagger_{4,0})|\text{vac}\rangle \nonumber \\
&= \frac{\sqrt{4!}}{\sqrt{4}^4}|D_4^{\{2,2\}}\>. 
\end{align}
with a success probability of $\frac{4!}{4^4}$.
From \cref{eq:NormalizedDicke}, we find the general success probability to be
\begin{align}
\label{eq:Psuc}
 P^{op}_N = \frac{ N!}{ N^N},
\end{align} which is always independent of $k$.

This is the success probability that can be achieved with linear transformation operators with vacuum when there is no interference at the detectors~\cite{Maser2010}. However, implementing the same operation in a unitary framework would require ancillary vacuum modes~\cite{Kasture2018}, thereby reducing the success probability due to additional unwanted interferences. A different solution would be to send all photons simultaneously into the same input mode of a unitary interference device, which implements the required transformation only for that mode~\cite{Kiesel2010}. We will discuss this scheme for qudit Dicke states using single-photon input states in the next section.

\subsection{Linear optical setup without ancillary photons}
\label{sec:QuditsWithoutAncilla}

In this section, we discuss an implementation of the linear transformation of \cref{no_ancilla_T} with unitary operators. The unitary operators are implemented  with symmetric multiport splitters,  
which transmit single input photons independently of the input mode to each output mode with the same probability.
The according unitary transformation is given by a discrete Fourier transform~\cite{Lim2005}
\begin{align}
	U_{jl} = \frac{1}{\sqrt{N}} \omega_{N}^{(j-1)(l-1)} ,
\label{eq:Fourier}
\end{align}
with $\omega_{N}=\exp{\left(\text{i} 2\pi/N\right)}$ being the $N$th root of unity.
Note that for $j=1$ or $l=1$, $U_{jl}=\tilde{T}_{jl}=1/\sqrt{N}$ [see \cref{no_ancilla_T}].

We first show that using a single symmetric multiport beam splitter yields the ideal success probability [see \cref{eq:Psuc}] when all required photons can be prepared in the same input mode deterministically.
We then include success probabilities for preparing $N$ photons in the same mode and demonstrate how this affects the overall success probability. Note that the following schemes solely rely on multiphoton interference and work for arbitrary photonic degrees of freedom that can encode qudits.

\begin{figure}
    \centering
    \includegraphics[width=1\linewidth]{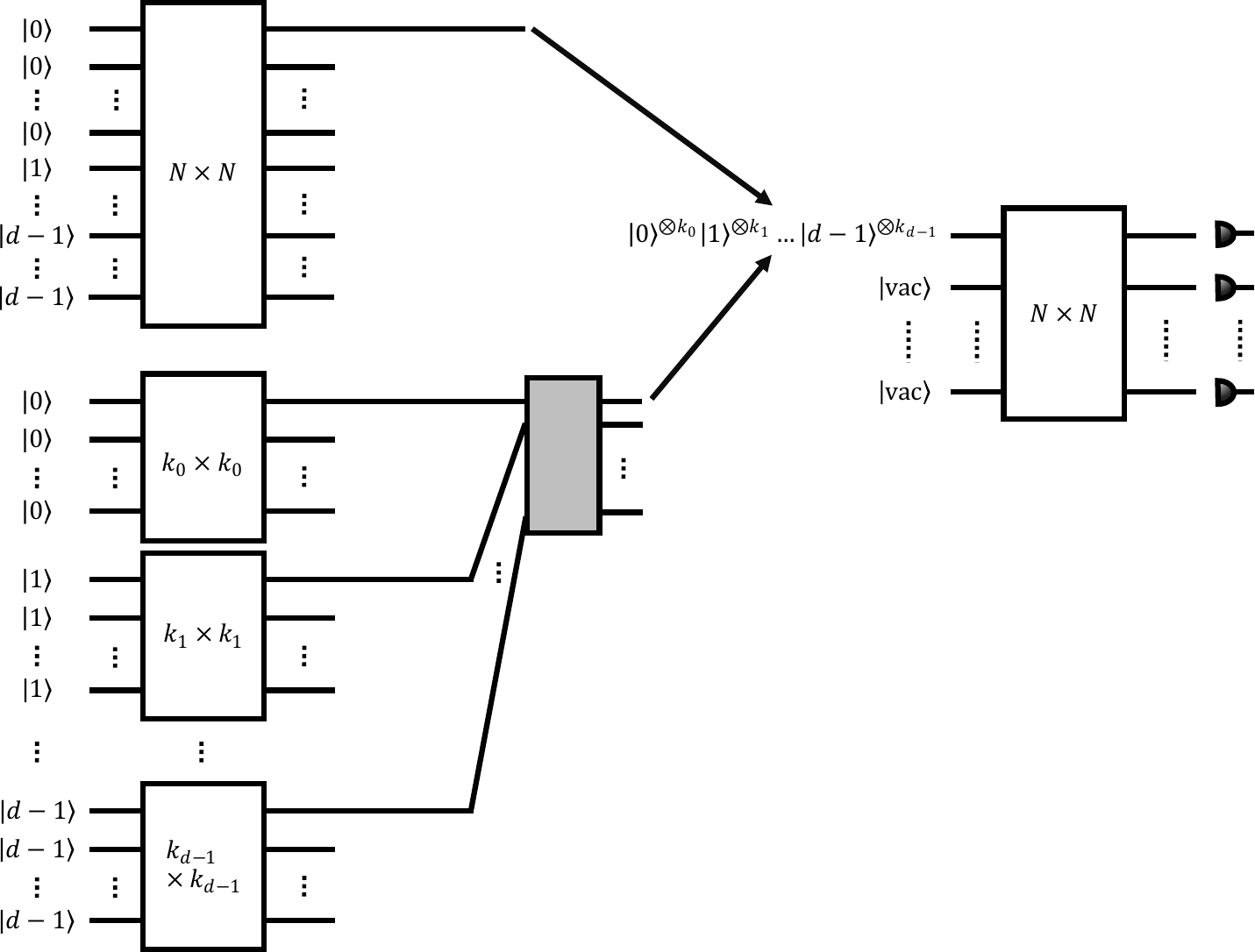}
    \caption{Right: Scheme to postselect qudit Dicke states from sending all photons into the same port of a symmetric multiport splitter. Left: Two schemes to prepare the input states, i.e., combine all photons in a single mode. In the upper scheme, all photons go into the same $N$-port splitter. In the lower scheme, photons of different qudit states $\ket{j}$ are sent into separate $k_j$-port splitters, respectively. The photons can then be collected in the same mode using a deterministic element.}
    \label{fig:SchemeHigherDimensional}
\end{figure}

Without loss of generality, we assume all photons entering the first input mode (see \cref{fig:SchemeHigherDimensional} right).
The input state is given by
\begin{align}
\label{eq:InputState}
	\ket{\Psi^{\{k\}}_{N,\text{in}}} = \prod_{j=0}^{d-1} \left( \frac{ (\ha_{1,j}^{\dagger})^{k_{j}}}{\sqrt{k_{j}!}} \right) \ket{\text{vac}}.
\end{align}
By utilizing the transformation relation [see \cref{eq:linear_evolution,eq:Fourier}] and setting $T_{1l} =U_{1l}=1/\sqrt{N} $, one obtains the output state
\begin{align}
	 \ket{\Psi^{\{k\}}_{N,\text{out}}}  & =  \frac{1}{\sqrt{N}^{N}}   \prod_{j=0}^{d-1} \left( \frac{\left(\sum_{l=1}^{N} \hb_{l,j}^{\dagger}\right)^{k_j}}{\sqrt{k_j!}}  \right)\ket{\text{vac}} .
\end{align}
One can directly see that this state is identical to \cref{qubit_operator_N}. Subsequently, the postselected state is identical to a qudit Dicke state and the success probability $p_{N,\{k\}}$ is identical to the upper bound $P_{N}^{op}$ [see \cref{eq:Psuc}]. The reason for this is that we have so far assumed to deterministically have access to arbitrary multiphoton Fock states. However, since the generation of multiphoton Fock states is challenging and limited to small photon numbers~\cite{Zhang2025PRR}, we discuss the generation of the input state given in \cref{eq:InputState} starting from single-photon input states.

One possibility would be to send the $N$ photons into the $N$ input modes of a symmetric $N$-port splitter separately (see \cref{fig:SchemeHigherDimensional} upper left). Due to the symmetry of the $N$-port splitter, all $N$ photons will go to the same output port with a probability of $p_{\{k\}\rightarrow 1}=(\prod_{j=0}^{d-1}k_{j}!)/N^N$. This reduces the overall success probability to
\begin{align}
\label{eq:Probability_Qudits_Realistic1}
	\tilde{p}_{N,\{k\}}  = \frac{N! \left(\prod_{j=0}^{d-1} k_{j}! \right)}{N^{2N}} N,
\end{align}
where the additional factor of $N$ arises due to the fact that one can add a symmetric $N$-port splitter to each of the input splitters' outputs and thereby perform the same postselection scheme $N$ times in parallel.
Note that this scheme has been investigated in Ref.~\cite{Kiesel2010} to generate qubit Dicke states using different input sources.

Alternatively, one could combine each group of $k_j$ photons in the state $\ket{j}$ using a separate $k_j$-port splitter  (see \cref{fig:SchemeHigherDimensional} lower left). For each splitter, the $\ket{j}$ photons will go to the same output mode with a probability of $p_{k_j\rightarrow 1}=k_j! /k_j^{k_j}$. Depending on the chosen encoding, one can combine the $d$ modes using a deterministic element and send all photons into the same input mode of the final $N$-port splitter.
Alternatively, one can send each group of $\ket{j}$ photons into a separate input mode of the same $N$-port splitter, leading to a Dicke state with relative phases. Both possibilities lead to the same overall success probability of
\begin{align}
\label{eq:Probability_Qudits_Realistic2}
	\tilde{p}_{N,\{k\}}'  = \frac{N!}{N^{N}} \left( \prod_{j=0}^{d-1} \frac{k_{j}!}{k_{j}^{k_{j}}} \right) k_{\text{min}} ,
\end{align}
with $k_{\text{min}}=\min\{k_0,k_1,\ldots,k_{d-1}\}$.  This factor arises due to the fact that one can perform one experiment for each of the $k_{\text{min}}$ outputs of the smallest $k_j$-port splitter in parallel. Note that \textcolor{black}{in the special case of $N=d$ and $k_0=k_1=\ldots=k_{d-1}=1$, \cref{{eq:Probability_Qudits_Realistic2}} becomes identical to the upper bound, i.e., \cref{eq:Psuc}.}

While sending all single photons into a single multiport splitter, in principle, requires only one additional optical element, using separate multiports, i.e., one per dimension, produces higher success probabilities (see \cref{fig:PlotCombined}). 

\section{Postselecting High-Dimensional Dicke States with Ancillary Photons}
\label{sec:SchemesQuditsAncilla}

The schemes in the previous section postselect $N$-partite Dicke states without using ancillary particles. In this section, we present an alternative scheme utilizing ancillary modes and particles to enhance the success probability.
We show that adding $K=k_1 + k_2 + \cdots + k_{d-1}=N-k_{0}$ ancillary photons allows for postselecting any symmetric Dicke state $\ket{D_{N}^{\{k\}}}$.

\subsection{Operator-level scheme}

To achieve our goal, we begin with an $N+K$ particle state
\begin{align}
|\Psi_{\text{in}}\> =
\prod_{i=1}^{N} \ha^\dagger_{i,0}\prod_{m_1=1}^{k_1}\hat{r}^\dagger_{m_1,1}\cdots\prod_{\substack{m_{d-1}\\=\sum_{h=1}^{d-2}k_h+1}}^{K}\hat{r}^\dagger_{m_{d-1},d-1}|\text{vac}\>,
\end{align} where $\ha^\dagger_{j,s}$ and $\hat{r}^\dagger_{j,s}$ denote particles in the $N$ main system modes and $K$ ancillary modes, respectively. 

Then we set the $(N+1) \times (N+K)$ 
matrix $T$ as follows:
\begin{align}\label{eq:k_ancilla_T}
T =
\begin{pmatrix}
\alpha\mathbb{I}_{N\times N} & \beta J_{ 1\times N} \\
\frac{1}{\sqrt{N}} J_{N\times K}   & O_{1\times K}
\end{pmatrix},~~~~~~~|\alpha|^2 + |\beta|^2 = 1,
\end{align} where $\mathbb{I}_{N\times N}$ is the identity matrix, $J_{p\times q}$ the all-one matrix, and $O_{p\times q}$ the all-zero matrix. With this transformation matrix, 
$N$ particles ($\hat{a}^\dagger_{i,0}$) are split into one output mode and a single main ancillary mode with equal probability amplitude each. Additionally, $K$ particles ($\hat{r}^\dagger_{m_{d-1},d-1}$) are split into the $N$ output modes symmetrically. Note that this operator preserves the permutation symmetry of Dicke states like~\cref{no_ancilla_T}. 

 Substituting $T$ into Eq.~\eqref{eq:N_transf}, the state is  transformed to
 \begin{align}\label{eq:transformed_ancilla}
    &\prod_{i=1}^N(\alpha\hb^\dagger_{i,0} + \beta \hat{s}^{\dagger}_{0})  \frac{1}{\sqrt{N}^K}\prod_{j=1}^{d-1}\frac{1}{\sqrt{k_j!}}(\sum_{m_j=1}^N \hb^\dagger_{m_j,j})^{k_j}|\text{vac}\>,
 \end{align} where $\hb^\dagger_{j,s}$ and $\hat{s}^\dagger_{s}$ denote particles in the $N$ output system modes and one ancillary mode respectively.
 
We postselect the cases when each mode in the main system receives exactly one particle and the ancillary mode receives $K$ particles. Then the final state is given by
\begin{align}\label{eq:postselected_ancilla}
 & \frac{\alpha^{N-K}\beta^K \big(\prod_{j=1}^{d-1}k_j!\big)}{\sqrt{N}^K\sqrt{\prod_{j=1}^{d-1}k_j!} } (\hat{s}^\dagger_{0})^K \nn \\
&\times \sum_{\sigma\in S_N}\Big( \prod_{j=1}^{N-K}\hb^\dagger_{\sigma(j),0}\prod_{l_1=N-K+1}^{N-K+k_1}\hb^\dagger_{\sigma(l_1),1} \nn \\
&~~~~~~~~~~~~~~~\cdots\times \prod_{l_1=N-k_{d-1}+1}^{N}\hb^\dagger_{\sigma(l_{d-1}),d-1} \Big)|\text{vac}\> \nn \\
&=  \frac{\alpha^{N-K}\beta^K\sqrt{\prod_{i=1}^{d-1}k_i!} 
\sqrt{N!K!}}{\sqrt{N}^K\sqrt{\prod_{i=0}^{d-1}k_i!}}\Big(\frac{(\hat{s}^\dagger_0)^K}{\sqrt{K!}}|\text{vac}\>_s\Big)\otimes |D_N^{\{k\}}\>.
\end{align}
Therefore, we obtain the product of the Dicke state and a K-particle state in the ancillary mode. By setting $|\alpha|^2=p$ and $|\beta|^2 = 1-p$, the success probability is given by 
\begin{align}\label{eq:succ_op_ancilla}
   P_{\{k\},\text{ancilla}}^{op} =\frac{N!K!}{(N-K)!} \frac{p^{N-K}(1-p)^K }{N^K}.
\end{align}
Linear optical implementation of this scheme in the next subsection will, again, decrease the success probability~\eqref{eq:succ_op_ancilla}.

\subsection{Linear optical setup with ancillary photons}

From Eq.~\eqref{eq:postselected_ancilla}, we can see that the transformation sends $K$ of the $\ket{0}$ photons into the ancillary modes and replaces them with the $k_j$ $\ket{j}$ photons for $j=1,\ldots,d-1$. In linear optical setups, this implies that $N-K$ $\ket{0}$ photons must be transmitted, while $K$ must be reflected.

\begin{figure}
	\centering
		\includegraphics[width=1.0\columnwidth]{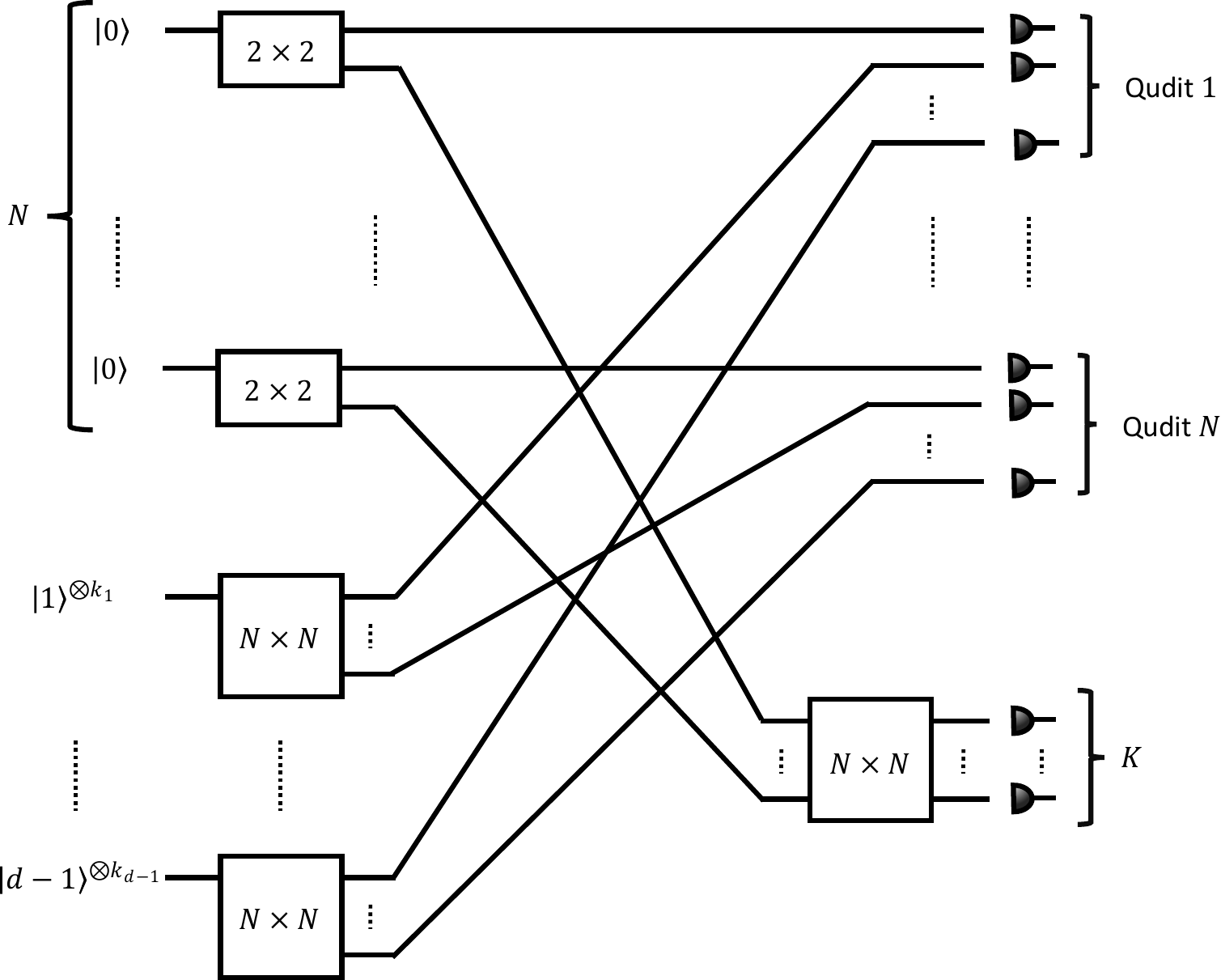}
	\caption{Setup to postselect qudit Dicke states with ancillary photons. We send in $N$ independent $\ket{0}$ photons into $N$ adjustable beam splitters. The $\ket{0}$ photons can either be transmitted to the output modes or reflected and detected at a symmetric $N$-port beam splitter. Simultaneously, we send $k_j$ $\ket{j}$ ($j=1,\ldots,d-1$) photons into symmetric $k_{j}$-port beam splitters, respectively, to replace the reflected $\ket{0}$ photons at the output modes. Finally, $N$-photon coincidences at the output modes in combination with $K=k_1 + \cdots + k_{d-1}$ reflected $\ket{0}$ photons postselect a symmetric Dicke state.}
	\label{fig:SchemeAncillasQudit}
\end{figure}

Our linear optical setup is shown in Fig.~\ref{fig:SchemeAncillasQudit}.
We send each of the $N$ $\ket{0}$ photons into an adjustable two-port beam splitter (transmissivity $p=|\alpha|^{2}$, reflectivity $1-p=|\beta|^{2}$).
We detect the reflected photons in the same output mode of a symmetric $N$-port splitter, erasing the which-path information and heralding the correct number of $\ket{0}$ photons.
For each dimension $j>0$, we insert $k_j$ photons in the state $\ket{j}^{\otimes k_j}$ into the first input mode of the $j$th $N$-port splitter and distribute them to the $N$ output qubits. We use path encoding, but other encodings can also be chosen. Finally, detecting for $N$-photon coincidences, i.e., one photon in each output qudit, allows for postselecting the Dicke state $\ket{D_{N}^{\{k\}}}$. In the following, we discuss the exact evolution of the quantum state and derive the overall success probability.

The complete input state is given by
\begin{align}
	\ket{\Psi_{\{k\},\text{in}}} = \left( \prod_{k=1}^{N} \ha_{k,0}^{\dagger} \right) \left(\prod_{j=1}^{d-1}\frac{\left( \ha_{N+j,j}^{\dagger} \right)^{k_j} }{\sqrt{k_j!}} \right) \ket{\text{vac}},
\end{align}
where the $N$ $\ket{0}$ photons are inserted into the first modes of the two-port splitters, while the $k_j$ $\ket{j}$ photons are inserted into mode $N+j$, i.e., the first mode of the $j$th $N$-port splitter.

The transformation of the two-mode beam splitters acting on the $\ket{0}$ photons is given by
\begin{align}
	\ha^{\dagger}_{k,0} \rightarrow \alpha \hb_{k,0}^{\dagger} + \beta \hs_{k,0}^{\dagger}  ,
\label{eq:TransformationDicke2}
\end{align}
where the transmitted photons go to the output modes $\hb_{k,0}^{\dagger}$, and the reflected photons go to the multiport modes $\hs_{k,0}^{\dagger}$ for heralding.
Further, each $\ket{j}$ photon inserted into mode $a_{N+j,j}$ is evenly distributed to one of the $N$ output modes [see \cref{eq:N_transf}]
\begin{align}
	\ha^{\dagger}_{N+j,j} \rightarrow \frac{1}{\sqrt{N}} \sum_{l=1}^{N} \hb_{l,j}^{\dagger} .
\label{eq:TransformationDicke1}
\end{align}

By using the transformation relations \cref{eq:TransformationDicke1,eq:TransformationDicke2}, one obtains the complete output state
\begin{align}
	&\ket{\Psi_{\{k\},\text{out}}} \nonumber \\
	&= \left( \prod_{k=1}^{N} \left[ \alpha \hb_{k,0}^{\dagger} + \beta \hs_{k,0}^{\dagger} \right] \right)  \left( \prod_{j=1}^{d-1} \frac{ \left( \sum_{l=1}^{N} \hb_{l,j}^{\dagger} \right)^{k_j} }{\sqrt{k_j!}\sqrt{N}^{k_j}} \right) \ket{\text{vac}}.
\end{align}
Comparing the transformed state to the operator description, we see that it is identical to \cref{eq:transformed_ancilla}, except for the fact that we now have $N$ distinct heralding modes $\hat{s}^{\dagger}_{k,0}$. As said before, we can erase this information by sending all heralding modes into an additional symmetric multiport splitter. We then have to detect all heralding photons in the same output mode of the multiport, reducing the success probability by $N/N^{K}$. We can therefore postselect a symmetric qudit Dicke state with a success probability of
\begin{align}
\label{eq:Probability_Qudits_Ancilla}
	p_{\{k\},\text{ancilla}} 
     = \frac{N! K!}{(N-K)!} \frac{p^{N-K} (1-p)^{K}}{N^{2K}} N.
\end{align}

Note that as long as one uses only a fixed output mode $l$ for heralding the Dicke state generation, e.g., $l=0$, the relative phases in the postselected Dicke state are stable. However, considering also $K$-photon coincidences in the other $(N-1)$ modes yields different relative phases. In these cases, feedforward unitary operations $U_{1},\ldots,U_{N}$  in the output modes $b_{1},\ldots,b_{N}$ can be used to eliminate the relative phases~\cite{Kim2020}. To this aim, we have to eliminate the phases introduced by the second $N$-port splitter, i.e., we choose a relative phase $\omega_{N}^{-(l_{1}-1)(l-1)}$ for $l_{1}=1,2,\ldots ,N$ for an $\ket{0}$ photon in output mode $l_{1}$.

Finally, one can maximize $p_{\{k\},\text{ancilla}}$ to find the best choice for $p$, i.e.,
\begin{align}
	p_{\text{max}} = \frac{N-K}{N} ,
\end{align}
which gives
\begin{align}
\label{eq:Probability_Qudits_Ancilla_max}
	p_{\{k\},\text{ancilla}}^{(\text{max})} =  \frac{N! K!}{(N-K)!} \frac{(N-K)^{N-K} K^{K}}{N^{N+2K}} N.
\end{align}

To account for single photon inputs, we add a factor of $k_i!/k_i^{k_i}$ for each dimension $i=1,\ldots,d-1$, which results in the final success probability
\begin{align}
\label{eq:Probability_Qudits_Ancilla_final}
	\tilde{p}_{\{k\},\text{ancilla}}^{(\text{max})} =  \frac{N! K!}{(N-K)!} \frac{(N-K)^{N-K} K^{K}}{N^{N+2K}} \left( \prod_{i=1}^{d-1} \frac{k_i !}{k_i^{k_i}} \right)  N.
\end{align}

Comparing this result to the linear optical setups for postselecting high-dimensional Dicke states without using ancillas [see \cref{eq:Psuc}], we observe that for $N \gg K$, \cref{eq:Psuc}/\cref{eq:Probability_Qudits_Ancilla_final} $\rightarrow 0$ for any choice of $k_i$ ($i=0,1,\ldots,d-1$). This shows that using ancillary photons will always boost success above the upper bound for the ancilla-free case. Intuitively, one can understand this boost by looking at the chosen optical elements. Replacing the full $N\times N$ multiport splitters with $2\times 2$ splitters for the photons in the state $\ket{0}$ reduces the number of possible output states these photons can produce and thereby increases the success probability. This becomes especially effective when the number of reflected photons $N-k_0$ is small. 
In the following section, we investigate this behavior in more detail for the two special cases of qubits and qutrits.

Note that for $d=2$ and $k_1=1$, our scheme and the calculated success probabilities become identical to the scheme for postselecting W states in Ref.~\cite{Kim2020}.

\section{Special Cases: Qubit and Qutrit Dicke States}
\label{sec:SpecialCases}

We now discuss our results in the special cases of photonic qubits and qutrits, which are relevant to current photonic implementations. 
We compare ancilla-free and ancilla-assisted schemes to verify which yields a higher success probability for given $N$ and $K$. 
For this, we present the respective success probabilities and compare the cases with and without ancillary photons.
Since high-fidelity multiphoton Fock states are still challenging to realize in current experiments~\cite{Zhang2025PRR}, we add \cref{eq:Psuc} as an upper bound for the ancilla-free schemes and not as a realistic scenario, showing that it can still be beaten using ancillary photons. Furthermore, we note that having access to deterministically generated multiphoton Fock states would also boost the schemes with ancillary photons, leading to an earlier advantage over the ideal ancilla-free scheme.

\subsection{Qubits}

The $N$ qubit Dicke state is defined as [see \cref{eq:DickeStateQudit}]
\begin{align}
\label{eq:DickeStateQubit}
    & \ket{D_{N}^{\{k_0, k_1\}}}  = \sqrt{\frac{k_{0}!k_{1}!}{N!}} \sum_{\sigma \in S_{N}^{\{k_0, k_1\}}} \ket{\underbrace{0\ldots0}_{k_{0}}\underbrace{1\ldots1}_{k_{1}}},
\end{align}
with $k_{0/1}$ photons in $\ket{0/1}$, $N=k_0+k_1$, and $K=k_1$.

\subsubsection{Without Ancillary Photons}

As discussed in \cref{sec:QuditsWithoutAncilla}, sending all photons into the same mode of a symmetric multiport beam splitter, one can postselect a symmetric qubit Dicke state with a success probability of $p_{N,\{k_0,k_1\}} = N!/N^{N}$ [see \cref{eq:Psuc}]. However, when starting from single-photon input states, this success probability will decrease.
Note that this case, i.e., generating symmetric qubit Dicke states using symmetric multiport beam splitters, has been discussed in Ref.~\cite{Kiesel2010}.

First, sending the $N$ photons into the $N$ input modes of a single symmetric $N$-port splitter (see \cref{fig:SchemeHigherDimensional} upper left) leads to [see \cref{eq:Probability_Qudits_Realistic1}]
\begin{align}
\label{eq:SinglePhotons_QubitDickeState_Probability1}
	\tilde{p}_{N,\{k_0,k_1\}} = \frac{N!k_0!k_1!}{N^{2N}}N.
\end{align}
Note that without the additional factor of $N$, this result is identical to the success probability of a similar scheme using a non-symmetric $2N$-port splitter~\cite{Kasture2018}.
Furthermore, in \cref{app:AlternativeQubitScheme} we discuss an alternative implementation using polarization encoding. It is an example of a no-touching scheme~\cite{Blasiak2019}, and requires a simple set of linear optical elements.

Second, sending the $k_0$ photons in the state $\ket{0}$ into the $k_0$ input modes of a $k_0$-port splitter, and the $k_1$ photons in the state $\ket{1}$ into the $k_1$ input modes of a $k_1$-port splitter, separately (see \cref{fig:SchemeHigherDimensional} lower left), leads to an overall success probability of [see \cref{eq:Probability_Qudits_Realistic2}]
\begin{align}\label{suc_pro_unit}
	\tilde{p}'_{N,\{k_0,k_1\}} = \frac{N!k_0!k_1!}{N^{N}k_0^{k_0}k_1^{k_1}} k_{\text{min}},
\end{align}
with $k_{\text{min}} = \min\{k_0,k_1\}$.

\subsubsection{With Ancillary Photons}
\label{sec:qubit_ancilla}

\begin{figure*}[t]
    \centering
    \includegraphics[width=1\linewidth]{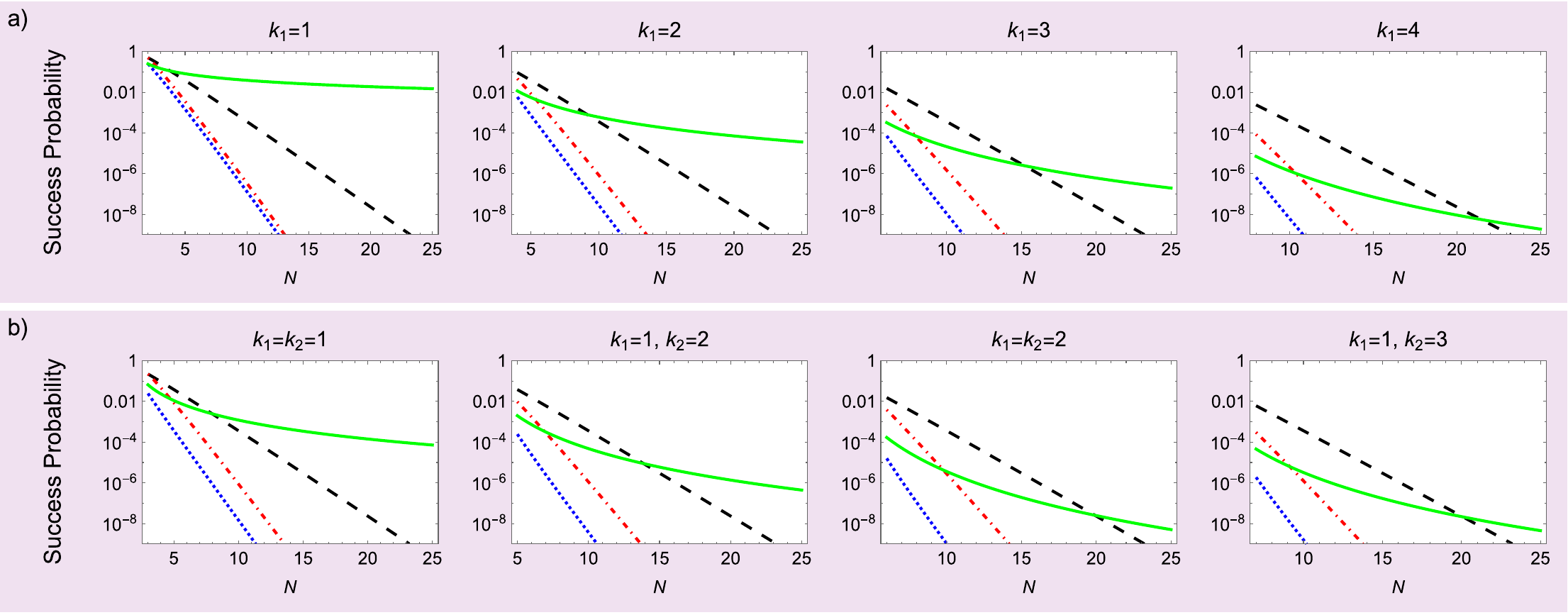}
    \caption{Maximum postselection probabilities for a) Qubits, and b) Qutrits. Black (dashed) lines: no ancillary photons, ideal case starting from multiphoton single-mode input state [a) and b) see \cref{eq:Psuc}]. Blue (dotted) line and red (dot-dashed) line: no ancillary photons, starting from single photon input states [a) see \cref{eq:SinglePhotons_QubitDickeState_Probability1} and \cref{suc_pro_unit} with additional factors of $N$ and $k_1$, respectively; b) see \cref{eq:Probability_Qutrits_Realistic1} and \cref{eq:Probability_Qutrits_Realistic2}]. Green (solid) line: ancillary photons [a) see \cref{eq:ptildemax}; b) see \cref{eq:ptildemaxQutrits}]. 
    The qubit results for $k_1=1$ correspond to the results of Ref.~\cite{Kim2020}.
  The success probabilities of all the cases in the plot remain above the per-window dark-count probability ($\sim 10^{-9}$) for modern superconducting nanowire single-photon detectors with dark-count rates around 1 count per second and nanosecond timing windows.}
    \label{fig:PlotCombined}
\end{figure*}

Starting from single-photon input states, the linear optical scheme with ancillary photons [see \cref{sec:SchemesQuditsAncilla}] yields the following success probability for qubits [see \cref{eq:Probability_Qudits_Ancilla_final}]:
\begin{align}
\label{eq:ptildemax}
	\tilde{p}_{\{k\},\text{ancilla}}^{(\text{max})} = \frac{N!k_1!}{k_0!} \frac{k_0^{k_0}k_1^{k_1} }{N^{N+2k_1} } \frac{k_1!}{k_1^{k_1}} N .
\end{align}

\subsubsection{Comparison}

\begin{figure*}[t]
\centering

\begin{minipage}{0.3\linewidth}
\centering
\begin{overpic}[width=\linewidth]{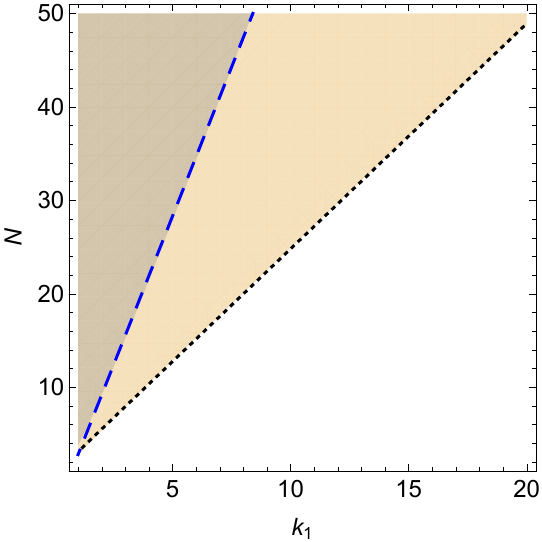}
\put(73, 18){{\fontfamily{lmss}\selectfont
\text{a) Qubits}}}
\end{overpic}
\end{minipage}
\hfill
\begin{minipage}{0.3\linewidth}
\centering
\begin{overpic}[width=\linewidth]{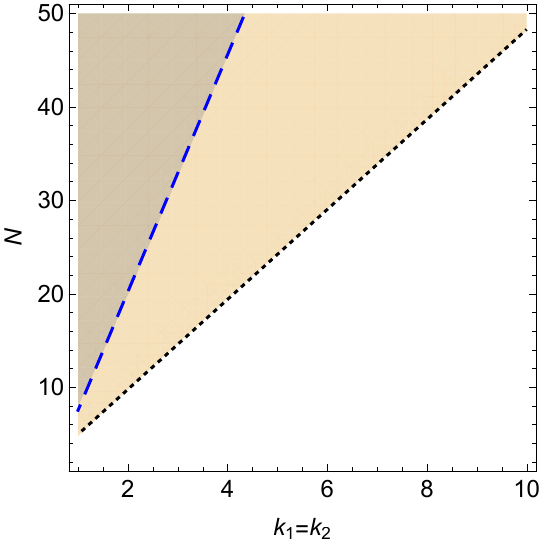}
\put(71, 18){{\fontfamily{lmss}\selectfont
\text{b) Qutrits}}}
\end{overpic}
\end{minipage}
\hfill
\renewcommand{\arraystretch}{1.5}
\begin{minipage}{0.3\linewidth}
\centering
\begin{tabular}{lcc}
\multicolumn{3}{c}{ {\fontfamily{lmss}\selectfont \text{c) Fit Results}} $f(x)=ax + b$} \\
\hline
 & $a$ & $b$ \\
\hline
{\fontfamily{lmss}\selectfont a)}  \text{$\cdots$} & $\num[round-mode=places, round-precision=3]{2.41059} \pm \num[round-mode=places, round-precision=3]{0.00468289} $ & $\num[round-mode=places, round-precision=3]{0.698508} \pm \num[round-mode=places, round-precision=3]{0.0560971}$ \\
{\fontfamily{lmss}\selectfont b)} \textbf{$\cdots$} & $\num[round-mode=places, round-precision=2]{4.8134} \pm \num[round-mode=places, round-precision=2]{0.0113897}$ & $\num[round-mode=places, round-precision=2]{0.148443} \pm \num[round-mode=places, round-precision=2]{0.0706711}$ \\
{\fontfamily{lmss}\selectfont a)}  \textcolor{blue}{- - -} & $\num[round-mode=places, round-precision=2]{6.37539} \pm \num[round-mode=places, round-precision=2]{0.055054}$ & $\num[round-mode=places, round-precision=2]{-3.57804} \pm \num[round-mode=places, round-precision=2]{0.341601}$ \\
{\fontfamily{lmss}\selectfont b)} \textcolor{blue}{- - -} & $\num[round-mode=places, round-precision=2]{12.6813} \pm \num[round-mode=places, round-precision=2]{0.132801}$ & $\num[round-mode=places, round-precision=2]{-5.14002} \pm \num[round-mode=places, round-precision=2]{0.440452}$ \\
\end{tabular}
\renewcommand{\arraystretch}{1.}
\end{minipage}
	\caption{Contour plots of the differences between the linear optical schemes without ancillary photons and the linear optical schemes with ancillary photons for a) postselecting a qubit Dicke state, and b) postselecting a qutrit Dicke state with $k_1=k_2$.
    Brighter and darker shaded area: success probability with ancillary photons exceeds the success probability without ancillary photons (single-photon input states), i.e., a)  \cref{eq:ptildemax} $>$ \cref{suc_pro_unit}, and b) \cref{eq:ptildemaxQutrits} $>$ \cref{eq:Probability_Qutrits_Realistic2}.
    Darker shaded area: success probability with ancillary photons exceeds the ideal success probability without ancillary photons (multiphoton Fock input states), i.e., a)  \cref{eq:ptildemax} $>$ \cref{eq:Psuc}, and b) \cref{eq:ptildemaxQutrits} $>$ \cref{eq:Psuc}.
    Dotted black line: Linear fit to the first contour line, i.e.,
    a)
    \cref{eq:ptildemax} $=$ \cref{suc_pro_unit}, and 
    b)
    \cref{eq:ptildemaxQutrits} $=$ \cref{eq:Probability_Qutrits_Realistic2}.
    Dashed blue line: Linear fit to the second contour line, i.e., 
    a) 
    \cref{eq:ptildemax} $=$ \cref{eq:Psuc}, and 
    b) 
    \cref{eq:ptildemaxQutrits} $=$ \cref{eq:Psuc}.
    We present all fit results in the table c) above.
    }
    \label{fig:ContourPlot}
\end{figure*}

To compare this result with the linear optical schemes without using ancillary photons [see \cref{eq:SinglePhotons_QubitDickeState_Probability1,suc_pro_unit}], we plot the respective probabilities for different $k_1\leq k_0$ in \cref{fig:PlotCombined}a).
Furthermore, we plot contour plots of the differences between the schemes in \cref{fig:ContourPlot}a) for up to $k_1=20$.
We find that for $N \lesssim (\num[round-mode=places, round-precision=3]{2.41059} \pm \num[round-mode=places, round-precision=3]{0.00468289} ) k_1 + (\num[round-mode=places, round-precision=3]{0.698508} \pm \num[round-mode=places, round-precision=3]{0.0560971})$ the ancilla-free scheme shows a higher success probability, and it will not be beneficial to use ancillary photons here. However, for larger $N$, one sees that using ancillary photons can be highly beneficial.
They even boost the success above the theoretical upper ancilla-free bound [see \cref{eq:Psuc}].
Finally, we note that in the regime where using ancillary photons becomes beneficial, the number of additional ancillary photons $k_1$ becomes small compared to $N$, and the additional cost of generating the ancillary photons will be small compared to generating the first $N$ photons.

\subsection{Qutrits}

The qutrit Dicke state is defined as
\begin{align}
\label{eq:DickeStateQutrit}
    & \ket{D_{N}^{\{k_0, k_1, k_2\}}}  \nonumber \\
    & = \sqrt{\frac{k_{0}!k_{1}!k_{2}!}{N!}} \sum_{\sigma \in S_{N}^{\{k_0, k_1, k_2\}}} \ket{\underbrace{0\ldots0}_{k_{0}}\underbrace{1\ldots1}_{k_{1}}\underbrace{2\ldots2}_{k_{2}}},
\end{align}
with $k_{0/1/2}$ photons in $\ket{0/1/2}$, $N=k_0+k_1+k_2$, and $K=k_1+k_2$.

\subsubsection{Without Ancillary Photons}

Again, by sending all $N$ photons into the same mode of a symmetric $N$-port splitter, one can postselect a symmetric qutrit Dicke state with a success probability of $p_{N,\{k_0,k_1,k_2\}} = N!/N^{N}$ [see \cref{eq:Psuc}]. The reduced success probabilities starting from single photon input states are given by [see \cref{eq:Probability_Qudits_Realistic1}]
\begin{align}
\label{eq:Probability_Qutrits_Realistic1}
	\tilde{p}_{N,\{k\}}  = \frac{N! k_{0}! k_{1}! k_{2}!}{N^{2N}} N,
\end{align}
when sending all single photons into the same $N$-port splitter, and [see \cref{eq:Probability_Qudits_Realistic2}]
\begin{align}
\label{eq:Probability_Qutrits_Realistic2}
	\tilde{p}_{N,\{k\}}'  = \frac{N!}{N^{N}} \left(  \frac{k_{0}! k_{1}! k_{2}!}{k_{0}^{k_{0}} k_{1}^{k_{1}} k_{2}^{k_{2}}} \right) k_{\text{min}} ,
\end{align}
with $k_{\text{min}} = \min\{k_0,k_1,k_2\}$, when sending all $k_{j}$ $\ket{j}$ photons into the same $k_j$-port splitter, respectively.

\subsubsection{With Ancillary Photons}

Starting from single-photon input states, the linear optical scheme for qutrits and with ancillary photons [see \cref{sec:SchemesQuditsAncilla}] gives [see \cref{eq:Probability_Qudits_Ancilla_final}]:
\begin{align}
\label{eq:ptildemaxQutrits}
	\tilde{p}_{\{k\},\text{ancilla}}^{(\text{max})} = \frac{N!(k_1+k_2)!}{k_0!} \frac{k_0^{k_0}(k_1+k_2)^{k_1+k_2} }{N^{N+2(k_1+k_2)} } \frac{k_1!k_2!}{k_1^{k_1}k_2^{k_2}} N .
\end{align}

\subsubsection{Comparison}

To compare this result with the schemes for postselecting qutrit Dicke states without ancillas, we plot the corresponding probabilities for different $k_1=k_2$ in \cref{fig:PlotCombined}b). Additionally, in \cref{fig:ContourPlot}b), we plot the contour plots of the differences of the different schemes for up to $k_1=k_2=10$.
Similarly to the qubit case, 
we find that for small $N \lesssim (\num[round-mode=places, round-precision=2]{4.8134} \pm \num[round-mode=places, round-precision=2]{0.0113897} ) k_1 + (\num[round-mode=places, round-precision=2]{0.148443} \pm \num[round-mode=places, round-precision=2]{0.0706711})$ the ancilla-free scheme shows a higher success probability and it will not be beneficial to use ancillary photons here. Again, going to a larger $N$ shows that using ancillary photons will be highly beneficial.
They even boost the success probability above the theoretical upper ancilla-free bound [see \cref{eq:Psuc}].
In total, in the regime where using ancillary photons is beneficial, the number of additional photons $k_1+k_2$ is small compared to $N$. In fact, we find the ratio of the total number of photons compared to ancillary photons is very similar in the qubit and in the qutrit case. This means that also in the qutrit case, the additional cost of generating the ancillary photons will be small compared to generating the first $N$ photons.

\section{Conclusion}
\label{sec:conclusion}

In this paper, we have presented a thorough discussion of linear optical setups for postselecting symmetric qudit Dicke states, with and without ancillary photons. All our schemes start from single-photon input states and are based on unitary operations only. We have analytically calculated the success probabilities for our schemes and compared the results.
Our analyses on qubit and qutrit cases demonstrate that ancilla-assisted schemes improve success probabilities for large $N$.

We hope that the presented schemes and the analytical discussion of their success probabilities will help to understand the actual cost of postselecting photonic qudit Dicke states. For this, it will be necessary to include different types of noise, e.g., photon loss, and realistic photon sources in the discussion.

Although we have been focusing on postselecting symmetric Dicke states, it would be interesting to adapt our schemes to postselect other types of Dicke states, such as phased and anti-symmetric Dicke states~\cite{Chiuri2010, Li2021, chin2024creating}.

Finally, it will be crucial to extend our schemes to heralding. Symmetric multiport splitters have already been used to herald high-dimensional GHZ states ~\cite{Chin2024,Bhatti2025} and qubit Dicke states~\cite{kang2025heralded}.
A straightforward way to enhance the presented schemes into heralding schemes would be to use non-destructive photon measurements in the output modes.

\section*{Acknowledgment}

D.B. thanks Stefanie Barz for helpful discussions.
This work was supported by the India-Japan Cooperative Science Programme under Grant 120257718, and the Japan Science and Technology Agency (JST) as part of Adopting Sustainable Partnerships for Innovative Research Ecosystem (ASPIRE), Grant Number JPMJAP24C1.

\appendix

\section{Alternative Qubit Scheme}
\label{app:AlternativeQubitScheme}

\begin{figure}[t]
	\centering
		\includegraphics[width=1\columnwidth]{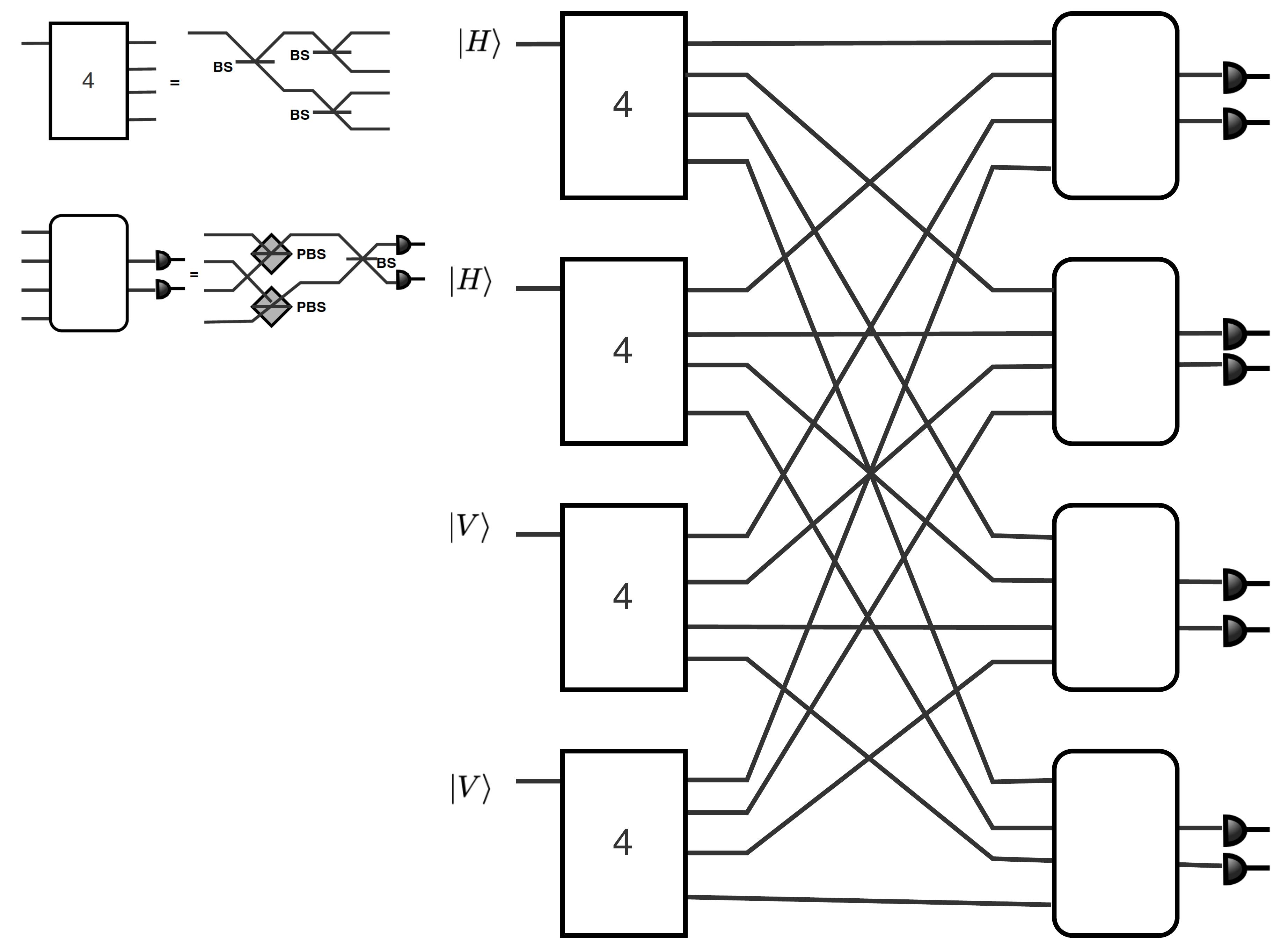}
	\caption{Linear optical scheme with only beam splitters (BSs) and polarizing beam splitters (PBSs). The 4-partite beam splitter is designed in the tree structure with three BSs (upper left), and we combine the photon path using two PBSs and one beam splitter (lower left).}
	\label{fig:dicke_linear_4}
\end{figure}

In this section, we present an alternative scheme that straightforwardly implements the linear transformation operator $\tilde{T}$ of Eq.~\eqref{no_ancilla_T} using only beam splitters (BSs) and polarized beam splitters (PBSs). From Fig.~\ref{fig:linear_transformation_no_ancillar}, a scheme in Fig.~\ref{fig:dicke_linear_4} is designed to generate $|D_4^{\{2,2\}}\>$. 
Four photons are prepared in different polarization input modes (horizontal $H$, vertical $V$); the initial state is given by
\begin{align}
    |\Psi_{\textrm{in}}\> = \ha^\dagger_{1,H}\ha^\dagger_{2,H}\ha^\dagger_{3,V}\ha^\dagger_{4,V}|\text{vac}\>.
\end{align} 
The photons are split equally into four different output modes through 4-partite splitters. The state evolves by rewiring as
\begin{align}
\frac{1}{\sqrt{4}^4}\big(\sum_{j=1}^4\hb^\dagger_{j,H}\big) \big(\sum_{k=1}^4\hb^\dagger_{k,H}\big) \big(\sum_{l=1}^4\hb^\dagger_{l,V}\big) \big(\sum_{m=1}^4\hb^\dagger_{m,V}\big)|\text{vac}\>.     
\end{align}
To minimize the interference, we design the detector as in Fig.~\ref{fig:dicke_linear_4}. Then, a quarter of the terms do not interfere with each other at the detectors. Therefore, by postselection, the final state is given by $|D_4^{\{2,2\}}\>$ with the success probability
$P_{suc}= \frac{4!}{4\cdot 4^4}$,  four times lower than the success probability of the operator-level scheme given in Eq.~\eqref{operator_final}. This probability is identical to the one in Eq.~\eqref{suc_pro_unit} when $(N,k)= (4,2)$. 

This scheme realizes the linear operation~\eqref{no_ancilla_T} using a relatively simple set of optical elements, i.e., a smaller set of BSs and PBSs. 
Another interesting feature of the scheme in Fig.~\ref{fig:dicke_linear_4} is that the photons do not touch each other before they arrive at the output modes, which can be considered the entanglement generation without touching~\cite{Blasiak2019}.
This circuit can be generalized to generate arbitrary Dicke states. The multipartite beam splitter can be implemented by recursively connecting beam splitters in the tree structure of BSs for any N, and the photon path can be combined by generalizing the $N=4$ case for even $N$, or by replacing it with symmetric multiport splitters.

\bibliography{Literature}{}

\end{document}